# When AI Classifies:
# What Counts as Public Administration?

**Shaoming Cheng**
**Department of Public Policy and Administration**
**Florida International University**
**scheng@fiu.edu**

**Laurie A. Schintler**
**Schar School of Policy and Government**
**George Mason University**
**lschintl@gmu.edu**

**Abstract:** This study examines how alternative systems of scholarly representation identify and characterize public administration (PA) and artificial intelligence-related PA (AI-in-PA) scholarship. The central question is not which system provides a definitive classification of PA, but how alternative systems of scholarly classification differ in identifying and representing its intellectual boundaries. With an exploratory and comparative analysis of Web of Science and OpenAlex, the study compares five approaches based on author-defined, citation-driven, and AI-assisted representations. The results highlight substantial differences and limited overlap in corpus size, publication type, DOI availability, publishing outlets, temporal patterns, and thematic structure. The alternative approaches tend to produce distinct representations because they identify substantially different bodies of scholarship rather than merely larger or smaller subsets of the same literature. The findings suggest that algorithmic knowledge organization influences how interdisciplinary scholarship is classified, connected, and understood and, epistemologically, how its visibility, structure, and intellectual identity are represented. AI-enabled scholarly classifications are not neutral but interpretative, potentially self-reinforcing, and may constrain the evolution and adaptation of disciplinary boundaries. Human expert judgment therefore remains essential and is complemented rather than replaced by AI-assisted scholarly classification.

# When AI Classifies:
# What Counts as Public Administration?

## Introduction

What counts as public administration (PA)? Traditionally, the boundaries of a discipline have been negotiated and defined collectively and collaboratively through practitioners and scholars, journals, citation communities, professional associations, intellectual traditions, and ongoing debates over the scope and identity of the field. Increasingly, however, these boundaries are mediated through scholarly indexing and classification systems assisted by artificial intelligence (AI). When bibliographic infrastructures rely more heavily on automated topic modeling, citation network analysis, semantic similarity measures, and algorithmic classification, AI is becoming an active actor, instead of merely technical tools, in the governance infrastructure of science. Algorithmic knowledge organization can influence which research themes appear central, peripheral, interdisciplinary within a field, as well as which scholarship becomes visible through scholarly discovery and retrieval systems. Epistemologically, what constitutes PA is no longer determined solely through intellectual developments within the discipline, but also by algorithms that classify, index, and retrieve disciplinary knowledge. They may impose computational representations of disciplinary boundaries and identity and may, potentially, introduce and reinforce systematic biases into how domains of knowledge are represented, connected, and understood.

PA has long been an interdisciplinary field, drawing theories, methods, and knowledge from political science, economics, management, sociology, information science, and public policy. Important developments relevant to PA, hence, often emerge through diverse publication channels and disciplinary communities (Olsen et al., 2026; Raadschelders, 2011). The rapid growth of AI-related scholarship further intensifies these interdisciplinary dynamics, as research rapidly combines administrative, ethical, managerial, and governance concerns with machine learning, natural language processing, computational methods, digital infrastructures, and emerging AI applications (McDonald et al., 2022; Zhu, 2026). AI intersects with PA in multiple ways: as a research method used to study administrative phenomena, as an object of scholarly inquiry concerning governance and public values, and as a technology applied within public and nonprofit organizations. Transdisciplinary research situated at the intersection of AI and PA (AI-in-PA) may be classified into adjacent domains such as computer science or data science despite its substantive relevance to PA. Because it combines substantive relevance to PA with theories, methods, and publication practices from multiple disciplines, AI-in-PA provides a particularly informative test case for examining how alternative systems of scholarly classification represent the visibility, or discoverability, and boundaries of interdisciplinary scholarship.

This study carries out a comparative analysis of Web of Science (WoS) and OpenAlex, with particular attention to OpenAlex's AI-assisted topic, concept, and subfield classification hierarchy. The central question is not which system provides a definitive classification of PA, but how alternative systems of scholarly classification differ in identifying and representing its intellectual boundaries. The comparison is analytically important because the two infrastructures rely on fundamentally different organizational logics. WoS is largely structured around human-curated, journal-level subject categories, and allows interdisciplinary studies to be retrieved through journal classifications and direct keyword searching. In contrast, OpenAlex relies more heavily on AI-assisted, publication-level classifications derived from citation networks, semantic similarity, topic assignment, and large language models. These differences may have important implications in visibility, representation, and ultimately the perceived boundaries of the field.

Methodologically, the study employs multiple complementary search and classification approaches across WoS and OpenAlex. To identify PA scholarship, the study utilizes the WoS

“Public Administration” category and the OpenAlex “Public Administration” subfield. Comparing WoS’s journal-based classification and OpenAlex’s publication-level, AI-assisted assignments provides insight into how different classification logics define and represent the broader PA scholarship. To identify AI-in-PA scholarship, both databases are searched using the layered AI framework developed in this study (Table 1), and the search terms are applied across five classification and metadata layers: (1) WoS titles, abstracts, and author-supplied keywords; (2) WoS Keywords Plus terms; (3) OpenAlex titles and abstracts; 4) OpenAlex Concepts; and (5) OpenAlex Keywords. These approaches represent three distinct systems of scholarly knowledge representation, specifically, author-defined, citation-driven, and AI-assisted, and show differences in corpus size, publication types, publishing outlets and overlap, temporal development, and thematic clustering and structure. Co-occurrence networks and ranked-term analyses are further used to compare how alternative representations portray the intellectual structure of AI-in-PA scholarship.

The significance of this study extends well beyond bibliometric comparisons of database coverage, indexing quality, or metadata completeness. It contributes to broader discussions of classification, representation, and boundaries by demonstrating that algorithmic knowledge organizations may produce substantially different understandings of what constitutes a field of scholarship. Existing discussions surrounding OpenAlex have focused primarily on openness, accessibility, metadata quality, and coverage relative to proprietary databases such as WoS and Scopus (Chawla, 2022). Comparatively less attention has been paid to the implications of AI-assisted classification systems for interdisciplinary domains where scholarly boundaries are fluid or even contested.

These epistemological implications are particularly important for PA because many contemporary policy challenges, including cybersecurity, climate change, public health, and AI itself, transcend traditional disciplinary boundaries. If interdisciplinary PA scholarship is systematically represented within adjacent fields rather than PA itself, its engagement with these issues may appear more limited than it actually is. As AI-assisted systems increasingly mediate scholarly discovery, classification, and evaluation, representation effects may influence not only how scholarship is organized but also what knowledge is discovered, cited, synthesized, taught, funded, and ultimately translated into policy and practice. They may also impact academic recognition, including hiring, promotion and tenure, and assessments of scholarly contribution, particularly for interdisciplinary research that does not fit neatly within established disciplinary categories. More fundamentally, the findings suggest that disciplinary boundaries are not rigid or static, but are continually negotiated, interpreted, and socially constructed through scholarly communities, publication practices, and systems of knowledge organization. As AI becomes embedded in the governance infrastructure of science through which scholarship is classified and discovered, algorithmic systems increasingly participate in this process. The challenge, therefore, is not simply whether AI classifies scholarship accurately, but how to balance the scalability and consistency of algorithmic knowledge organization with the inherently dynamic, contested, and socially constructed nature of disciplinary knowledge.

## Evolution of PA's Disciplinary Boundaries

What counts as PA has long been a subject of debate. Unlike disciplines organized around relatively established theoretical or methodological traditions, PA has developed primarily through its concern with government, governance, public organizations, and administrative problems. Its intellectual identity has consequently been shaped by continuing debates over the boundaries of administrative knowledge and its relationship with neighboring fields. Raadschelders (2011) characterizes PA as an interdisciplinary study of government, emphasizing the diverse intellectual traditions contributing to its development. Kettl (2000) similarly documents the field's recurring struggles over its identity, relevance, and theoretical foundations. These debates reflect a persistent tension between establishing a recognizable disciplinary identity and maintaining the intellectual openness required to examine complex administrative and societal problems.

The historical development of PA illustrates this tension. Wilson's (1887) *The Study of Administration* and Goodnow's (1900) subsequent work helped establish administration as a distinct subject of systematic inquiry while situating it within broader questions of politics and government. The politics–administration dichotomy became an influential framework for defining the field, although its interpretation and practical implications have been contested. As Kettl (2000) explains, PA initially occupied a central position within political science but gradually experienced increasing intellectual and institutional separation. The rise of behavioralism and changing expectations for theoretical and methodological rigor challenged established administrative approaches, while competing perspectives on management, bureaucracy, and governance further diversified the field. Frederickson (1999) likewise addresses the changing position of PA within political science and its efforts to establish a distinct intellectual identity. Rather than converging around a unified theoretical paradigm, PA developed through multiple approaches to understanding administrative institutions, political authority, organizational behavior, and public management (Kettl, 2000; Riccucci, 2010).

This intellectual diversity has been accompanied by evolving relationships with other disciplines. Through a scientometric analysis of 66 journals, Wagner and Raadschelders (2025) trace PA's development from early multidisciplinary engagement through disciplinary consolidation toward a more mature interdisciplinary field. Their findings suggest that PA has developed a recognizable intellectual identity while continuing to draw upon and contribute to neighboring disciplines. Ni, Sugimoto, and Robbin (2017) similarly examine the evolution of PA through the publication and citation patterns of *Public Administration Review*, documenting the journal's role in connecting PA with political science and other fields.

Yet interdisciplinary engagement does not necessarily produce integrated scholarly communities. Vogel (2014) finds that PA and organization studies, despite their shared intellectual roots, have developed relatively distinct publication and citation communities, with knowledge transfer occurring primarily from organization studies to PA. Han et al. (2020) highlight the limited integration of macroeconomic perspectives into PA scholarship and argue for stronger engagement between the two fields. Olsen, Bendtsen, and van Leeuwen (2026) identify a substantial body of PA-relevant research published in economics journals that is largely disconnected from mainstream PA citation networks. Beyond disciplinary relationships, Frank and Kunz (2025) emphasize the need to reconnect PA education, research, and practice, drawing attention to the institutional and professional dimensions of the field's intellectual development.

This distinction raises a further question about how disciplinary boundaries are identified and represented. Studies of scholarly communication demonstrate that disciplinary relationships can vary depending on the publications, journals, and citation connections used to characterize a field (Ni et al., 2017; Vogel, 2014; Wagner & Raadschelders, 2025). Examining regional science, Chen and Schintler (2023) similarly demonstrate how scientometric approaches can be used to investigate a field's evolving intellectual position within the broader scientific landscape. Moreover, the identification of a scholarly field depends partly on the bibliographic infrastructure through which publications are indexed and classified. Chavarro, Ràfols, and Tang (2018) demonstrate that inclusion in the WoS reflects selective indexing practices that should not be treated as a straightforward indicator of scholarly quality. The development of alternative scholarly databases, including OpenAlex, has expanded the possibilities for examining scientific production beyond established commercial indexing systems (Chawla, 2022). These differences matter for interdisciplinary fields such as PA because journal-based, publication-level, and algorithmically generated classifications may identify different populations of relevant scholarship. Consequently, the boundaries represented in bibliometric analyses need not coincide with the substantive boundaries of the field itself.

The emergence of AI makes these questions particularly salient. AI has become a subject of PA research, a methodological resource, and an instrument of administrative practice. McDonald et al. (2022) identify AI and related technological developments among the emerging directions of PA research, while Wang, Yin, and Hu (2026) examine AI's applications and governance challenges in public administration. Zhu (2026) conceptualizes the field's relationship with AI as developing “with, of, and through AI,” encompassing AI-enabled administrative activities, the governance of AI, and its implications for the production of administrative knowledge. AI-related PA scholarship therefore extends across research on public-sector applications, governance and accountability, and the use of AI in research and professional education (Li et al., 2025; Shark, 2024). These dimensions connect PA with computer science, information science, management, economics, law, and public policy, making the identification of AI-related PA research particularly sensitive to disciplinary and topical classifications.

At the same time, AI is increasingly reshaping scientific research itself. Gao and Wang (2024) examine the use and potential benefits of AI across scientific research, while Bianchini et al. (2026) investigate where and when AI contributes to scientific outcomes. Hao et al. (2026) further examine the relationship between AI tools, scientific impact, and the concentration of research attention. Although these studies provide important evidence about AI's role in knowledge production, their objectives differ from examining how alternative bibliographic systems classify AI-related scholarship within a particular interdisciplinary field. Their findings therefore motivate, rather than resolve, questions about how the choice of database and classification approach shapes the research populations and intellectual structures observed in bibliometric analyses.

For PA, these questions are especially consequential given the field's historically negotiated disciplinary identity and its extensive intellectual connections beyond conventional PA journals. The same research on AI-enabled public services, algorithmic accountability, administrative decision-making, or digital governance may be identified differently depending on whether PA membership and AI relevance are determined through journal classifications, author-supplied terminology, citation relationships, or AI-assisted publication-level classifications. Existing scholarship establishes that PA's substantive and publication boundaries do not fully coincide, but it does not establish whether different classification approaches produce comparable representations of AI-related PA research. Addressing this question requires examining classification itself rather than treating the resulting bibliographic corpus as a definitive representation of the field. By comparing WoS and OpenAlex and their alternative approaches to identifying AI-related scholarship, the present study investigates how classification choices influence what is recognized and represented as AI research in PA.

## Research Questions

This study adopts an exploratory approach and examines how alternative systems of scholarly classification represent PA and AI-in-PA scholarship. The objective is not to explain why one system causes a particular outcome, but rather to examine how alternative systems differ in identifying, organizing, and representing the same body of scholarship. It addresses three research questions.

**Research Question (RQ) 1:** How do WoS and OpenAlex differ in their representation of the broader PA literature?

**RQ 2:** How do alternative representations identify AI-in-PA scholarship?

**RQ 3:** How do alternative representations characterize the thematic structure of AI-in-PA scholarship?

## Bibliographic Data Collection and Search Strategy

### *Operationalization of PA*

This study draws on bibliographic records from WoS and OpenAlex. WoS is a widely used source of high-quality scholarly publications with strong coverage of established journals but more limited coverage of emerging, regional, and non-traditional outlets (Chavarro et al., 2018). OpenAlex (https://openalex.org), named after the ancient Library of Alexandria in Egypt, is an open access bibliographic catalog containing more than 200 million scholarly works and related research outputs (Chawla, 2022). Developed as the successor to Microsoft Academic Graph, OpenAlex aggregates metadata from journals, books, conference proceedings, repositories, preprints, dissertations, and other scholarly sources. Compared with WoS, OpenAlex provides broader and more inclusive coverage of scholarly communications, particularly beyond conventional journal literature. However, this broader coverage may also introduce greater topical noise and metadata variability and requires additional filtering and interpretation. To maintain comparability across the two databases, the analysis is restricted to English-language publications.

The analysis proceeds in two stages. First, it identifies the broader PA literature using each database's disciplinary classification system. Second, it identifies AI-in-PA scholarship within those corpora using alternative search and metadata representations. The study focuses on publications at the intersection of AI and PA. To identify PA scholarship in WoS, we use the "Public Administration" subject category. This approach captures journals, conference proceedings, and books classified as PA but may exclude relevant work published in neighboring fields such as political science, public policy, development studies, or urban planning. We do not supplement the search with additional categories or PA-specific keywords because doing so could bias the thematic composition of the resulting corpus.

OpenAlex organizes scholarly knowledge through a hierarchical classification system in which "Topics" represent the most granular level, aggregating into broader "Subfields," which then roll up into larger "Fields," and ultimately into highly aggregated "Domains." Within this structure, "Topics" are not manually curated but are generated through a large-scale automated clustering system grounded in citation-network analysis. Once a cluster is formed, OpenAlex assigns a human-readable label by extracting the titles and abstracts of the 100 most representative papers and submitting them to a GPT-based large language model with a prompt requesting the most accurate name for that specific research area. To operationalize PA within OpenAlex, a publication is classified as PA when it is assigned to a Topic belonging to the "Public Administration" Subfield (3321). Because OpenAlex classifications are algorithmically generated and organized around topic and subfield structures, some interdisciplinary or boundary-spanning PA scholarship may not be fully captured within this classification approach, particularly in emerging areas such as AI-related research.

### *AI Search Keywords and Strategy*

To organize the heterogeneous set of AI-related methods identified in the keyword search, this study adopts a layered conceptual framework consisting of four levels: foundation, architecture, model, and action. Table 1 summarizes the categories of AI techniques and associated keywords included in our search. The distinction between foundation, architecture, model, and action layers reflects how AI knowledge is organized and discussed, from general learning principles to system design, model instantiation, and downstream use. Such a layered framework allows AI-related concepts to be grouped by function rather than by field or application, and enables systematic comparison across otherwise incomparable research outputs.

[Insert Table 1 here]

The framework also prioritizes broad AI paradigms and widely recognized conceptual categories that structure contemporary AI discourse across fields. Accordingly, highly specific algorithms or techniques, such as random forest, support vector machines, or neural networks, are not included as primary search terms because they represent specific methodological implementations within broader AI domains. Including large numbers of individual algorithms would substantially expand and fragment the search strategy but offers limited additional conceptual distinction at the field level.

***Identification of AI-in-PA Scholarship***

The AI search terms presented in Table 1 are applied across five classification and metadata layers: (1) WoS titles, abstracts, and author-supplied keywords; (2) WoS Keywords Plus terms; (3) OpenAlex titles and abstracts; (4) OpenAlex Concepts; and (5) OpenAlex Keywords. As summarized in Table 2, these approaches represent three distinct systems of scholarly knowledge representation: author-defined, citation-driven, and AI-assisted.

[Insert Table 2 here]

The author-defined approaches rely on information provided directly by authors. The WoS Author approach identifies AI-in-PA scholarship using titles, abstracts, and author-supplied keywords, whereas OA Author relies on titles and abstracts because author-supplied keywords are not available in OpenAlex. The citation-driven approach, WoS Keywords Plus, identifies publications using algorithmically generated terms derived from cited references. The AI-assisted approaches, OA Concepts and OA Keywords, rely on algorithmically assigned concepts and keywords generated through OpenAlex's semantic classification system.

For OA Concepts and OA Keywords, AI-related publications are identified using the AI dictionary derived from the layered AI framework presented in Table 1. OpenAlex assigns relevance scores indicating the strength of association between a publication and a concept or keyword. A sensitivity analysis of thresholds ranging from 0.1 to 0.9 resulted in the selection of a 0.20 cutoff, which balances coverage and topical precision. Publications are retained only when at least one AI-related concept or keyword meets or exceeds this threshold.

***Term Construction and Co-Occurrence Networks***

To examine the conceptual structure of AI-in-PA scholarship, we construct co-occurrence networks in which terms are linked based on their co-occurrence within individual publications. Node size reflects term prominence, while edges represent the frequency with which terms appear together. Two complementary representations are analyzed. First, a co-occurrence network is constructed from WoS author-supplied keywords. Because these terms are selected by authors, they provide an author-defined representation of the literature and reflect how researchers intentionally frame and position their work.

Second, to establish a common basis for comparison across databases, co-occurrence networks are constructed from n-grams extracted from titles and abstracts in both WoS and OpenAlex. Unigrams, bigrams, and trigrams are extracted to capture recurring concepts and multiword expressions such as “machine learning,” “natural language processing,” and “large language models.” Similar n-gram approaches have been increasingly used in bibliometric and computational text analyses of AI-related research (e.g., Gao & Wang, 2024). Unlike author-supplied keywords, which reflect authors’ intentional conceptual framing, title and abstract n-grams provide a broader representation of publication content.

Separate networks are therefore constructed for WoS author keywords, WoS title/abstract n-grams, and OpenAlex title/abstract n-grams. In all networks, edges are defined by document-level co-occurrence and weighted by co-occurrence frequency. Terms appearing fewer than five times are excluded from the WoS keyword network, and the n-gram networks are limited to the 100 most frequent terms to facilitate comparability across databases. Network visualization and community detection are performed using VOSviewer and the Louvain algorithm. The analysis is descriptive and comparative, focusing on observed differences among the retrieved bibliographic corpora rather than making statistical inferences about a broader population of scholarship.

## Bibliometric Findings

### *PA Representation in WoS and OpenAlex*

Searches were conducted separately in April 2026 in WoS and OpenAlex using the WoS category of "Public Administration" and OpenAlex's Subfield 3321 "Public Administration" for the period 1980-2026. Appendix 1 presents the entire WoS search query, while Appendix 2 documents the OpenAlex API retrieval, data-processing, and keyword-filtering procedures. These searches aim to capture the full corpus of PA-related scholarship, from which AI-in-PA studies are identified. The WoS search returned 160,354 PA-related records which decreased to 152,697 records when only English-language publications are included. In OpenAlex, the search returned 280,213 PA-related records, of which 246,246 were English-language publications.

[Insert Figure 1 here]

Figure 1 compares the publication types of PA records indexed in WoS and OpenAlex. Despite differences in database size, the overall publication profiles are broadly similar. Articles constitute the largest category in both databases, followed by book chapters and books. PA scholarship remains centered on journal-based research and maintains a substantial book-oriented tradition. OpenAlex exhibits substantially broader coverage across publication types. It contains more records than WoS in every major category and includes publication types that are largely absent from WoS, most notably preprints and dissertations. Some of these differences reflect alternative classification practices. For example, OpenAlex frequently incorporates conference proceedings within its article category, whereas WoS records proceedings and Early Access items separately. Even after accounting for these differences, OpenAlex provides a substantially broader representation of PA scholarship.

The inclusion of preprints and dissertations is particularly noteworthy because these publication types often capture emerging research before it appears in the peer-reviewed literature. Consequently, OpenAlex is more likely to identify developing topics, interdisciplinary collaborations, and other boundary-spanning scholarship that may not yet be visible in WoS. Overall, OpenAlex provides a broader and more diverse representation of PA scholarship than WoS. These differences may extend beyond database size but reflect alternative approaches to scholarly classification and knowledge organization.

### ***AI-in-PA Representation and Comparison***

Corpus Sizes. The five representations identify markedly different AI-in-PA corpora (Figure 2). The WoS Author representation yields the largest corpus (959 publications), nearly twice the size of the OA Concepts corpus (494 publications) and substantially more than OA Keywords (387 publications), OA Author (204 publications), and WoS Keywords Plus (133 publications). Despite OpenAlex's broader and more inclusive coverage of the overall PA literature, its AI-assisted representations identify substantially fewer AI-in-PA publications than the WoS Author approach. Broader database coverage itself does not automatically translate into a larger AI-in-PA corpus. Instead, systems of scholarly representation and retrieval play an important role in determining which publications become visible as AI-in-PA scholarship.

[Insert Figure 2 here]

Publication Types. The distributions of publication types shown in Figure 3 reveal important differences across the five AI-in-PA representations. Although journal articles constitute the largest publication type in all representations, the two WoS representations are disproportionately concentrated in journal articles and conference proceedings. The WoS Author representation, in particular, contains more than 300 proceedings papers, which is consistent with the rapid development of AI-related research, where conference venues often serve as important outlets for the dissemination of new methods, applications, and findings. In contrast, the OpenAlex representations capture a broader mix of publication types. In addition to journal articles, OA Concepts and OA Keywords identify substantial numbers of books, book chapters, and preprints, while the OA Author representation exhibits a similar pattern on a smaller scale. Compared with the WoS representations, OpenAlex identifies AI-in-PA scholarship across a more diverse range of publication formats. These findings indicate that differences among representations extend beyond corpus size to corpus composition. This contrast is especially notable when compared with the broader PA literature (Figure 1), where WoS and OpenAlex exhibit broadly similar publication profiles. Within AI-in-PA, however, the representations diverge considerably, suggesting that differences in scholarly representation become more pronounced for interdisciplinary and emerging research domains, where disciplinary boundaries are fluid and rapidly evolving.

[Insert Figure 3 here]

Availability and Overlap of Digital Object Identifier (DOI). Table 3 reports DOI presence and DOI overlap across the five representations. DOI availability is highest for WoS Author (96.3%), WoS Keywords Plus (94.0%), and OA Author (92.7%). In contrast, OA Concepts (57.1%) and OA Keywords (60.7%) exhibit substantially lower DOI presence. DOI provides a common identifier for comparing publications across representations. All 235 DOI-linked publications identified by OA Keywords also appear in OA Concepts, accounting for 83.3% of the DOI-linked OA Concepts corpus. OA Author, however, shares only 34 publications with OA Concepts and 19 with OA Keywords, thus suggesting limited commonality between OpenAlex's author-defined and AI-assisted representations.

[Insert Table 3 here]

Cross-database overlapping is even more limited. WoS Author shares only two DOI-linked publications with OA Concepts and two with OA Keywords, while WoS Keywords Plus shares none with either OpenAlex AI-assisted representation. Furthermore, the two WoS representations overlap by only 79 publications, representing 8.6% of the DOI-linked WoS Author corpus. These findings demonstrate that the five representations, even when derived from the same database, produce substantially different representations of AI-in-PA scholarship rather than simply identifying larger or smaller subsets of the same literature. The near absence of overlap between the OpenAlex AI-assisted and WoS representations is particularly striking. Because these comparisons are restricted to records with identifiable DOIs, the observed overlap may not fully represent the relationships among the complete corpora, particularly for the OpenAlex Concepts and Keywords approaches.

<u>Publication Outlets</u>. Publication outlets provide additional evidence that alternative representations position AI-in-PA scholarship within different scholarly and professional communities. Figure 4 compares the top 15 publishing outlets identified by four of the five approaches. The WoS Keywords Plus representation is omitted because it identifies the smallest corpus and substantially overlaps with the WoS Author representation. The WoS Author representation is concentrated in established PA, public policy, and digital government journals and conference proceedings, including *Data & Policy*, *Public Management Review*, *Public Administration Review*, *Public Administration*, *Regulation & Governance*, *Journal of European*

*Public Policy*, and major digital government conferences. This profile reflects a relatively bounded and governance-oriented representation of AI-in-PA scholarship. In contrast, the OA Author representation exhibits only a limited presence in traditional PA journals, but a concentration in social work journals, including *Social Work Education*, *Journal of Social Work*, *The British Journal of Social Work*, *European Journal of Social Work*, and *Administration in Social Work*. The OpenAlex AI-assisted representations, OA Concepts and OA Keywords, exhibit highly similar profiles and are dominated by repositories, institutional platforms, ebook collections, and interdisciplinary or technical venues, such as eCommons, Zenodo, and *IEEE Transactions on Pattern Analysis and Machine Intelligence*.

[Insert Figure 4 here]

The analysis of outlet overlapping further reinforces these differences. To facilitate interpretation, Figure 5 presents a Venn diagram comparing three alternative representations. The OA Keywords representation is omitted because it is entirely subsumed within OA Concepts, and including a fourth representation would add visual complexity without much added insight. Among the top 20 outlets of each approach, there is no single outlet, journal or alike, shared across all three approaches. WoS Author shares only one outlet with OA Author and none with OA Concepts, whereas OA Author and OA Concepts share four outlets. A similar pattern emerges when total numbers of publications are considered. There is no single publication, article, proceeding, or alike, shared by all three approaches and the overlapping is concentrated primarily within the OpenAlex representations. These findings suggest that systems of scholarly knowledge classification are not neutral. Rather, they influence which scholarly communities become visible, how knowledge is communicated and discovered, and how the intellectual boundaries and identity of an interdisciplinary field are represented.

[Insert Figure 5 here]

Temporal Development. The cumulative publication trajectories shown in Figure 6 illustrate significantly different temporal narratives of AI-in-PA scholarship. The OpenAlex AI-assisted representations, OA Concepts and OA Keywords, identify publications dating back to 1980 and exhibit relatively steady, almost linear growth over four decades. This may suggest the gradual diffusion of AI-related concepts and methods, originating outside of the PA literature, into PA through interdisciplinary scholarship. In contrast, publications of the WoS approaches are concentrated in recent years, with particularly rapid, almost exponential growth beginning around 2023. This sharp increase coincides with the widespread adoption of generative AI following the public release of ChatGPT and suggests that AI has recently become a much more visible area of inquiry within the PA literature.

[Insert Figure 6 here]

The summary statistics in Table 4 reinforce these differences. The OpenAlex AI-assisted representations are centered approximately a decade earlier than the WoS-based representations, with mean publication years around 2010 and median publication years of 2011-2012. In contrast, the WoS representations are concentrated almost entirely in the recent surge of AI-related research, with mean publication years after 2022 and median publication years of 2024. The OA Author representation falls between these two patterns, with a mean publication year of 2020 and a median of 2023.

[Insert Table 4 here]

The contrasting temporal patterns likely reflect differences between general-purpose and field-specific systems of scholarly representation. OpenAlex is designed, probably optimized, to organize scholarship across general, broader scientific literature, where AI has long been applied and researched. But the WoS Author representation reflects how the PA research community specifically identifies and labels its own scholarship. These different classification logics may produce contrasting temporal patterns shown in Figure 6. The OpenAlex AI-assisted representations emphasize the broader conceptual diffusion and integration of AI into PA through interdisciplinary scholarship, while the WoS Author representation emphasizes the more recent visibility of AI as an explicitly identified research domain within the PA community. These temporal patterns reflect differences in how the five approaches identify and represent AI-in-PA scholarship and should not be interpreted as definitive evidence of the field's actual historical development.

### *Representations and Comparison of Thematic Clusters*

The conceptual structures of AI-in-PA are compared with three alternative representations: WoS author-supplied keywords, WoS title- and abstract-derived n-grams, and OpenAlex title- and abstract-derived n-grams. To facilitate comparison, the ranked terms in Table 5 are grouped into three broad domains: (1) AI Methods and Technologies, (2) Governance, Policy, and Administration, and (3) Societal, Organizational, and Professional Contexts. These categories serve as analytical lenses rather than mutually exclusive classifications.

[Insert Table 5 here]

The ranked terms suggest clear differences in thematic emphasis across representations. The WoS Author representation is dominated by AI methods and technologies, including artificial intelligence, machine learning, natural language processing, generative AI, and large language models, alongside governance concepts such as public sector, governance, regulation, and accountability. In contrast, the WoS title- and abstract-derived n-grams shift emphasis toward governance and administrative processes, with concepts such as public sector, public administration, regulation, accountability, and administrative appearing more prominently than specific AI techniques. The OpenAlex title- and abstract-derived n-grams exhibit the broadest contextual orientation, giving greater prominence to policy, practice, education, professional, social workers, ethics, and technology. Compared with the WoS representations, OpenAlex identifies a wider range of professional and societal contexts in which AI is discussed.

[Insert Figure 7 here]

Figure 7 illustrates how prominent concepts are organized within each representation. Conceptual prominence does not necessarily correspond to structural importance: highly ranked terms may play relatively peripheral roles in the network, whereas less frequent terms may connect multiple thematic areas. The WoS Author network exhibits the strongest AI centrality. AI serves as the dominant hub linking machine learning, natural language processing, generative AI, governance, regulation, and public-sector applications, producing a relatively coherent AI-centered structure. In contrast, the WoS title- and abstract-derived network is organized more around governance and administrative processes than around AI methods. Although AI remains prominent, governance, public sector, accountability, regulation, and institutional concepts occupy more central positions, reflecting broader substantive contents and themes. The OpenAlex title- and abstract-derived network displays the broadest contextual structure. Rather than centering primarily on AI methods or governance, it emphasizes professional practice, education, ethics, implementation, and societal applications. AI remains an important concept, but it functions within a wider interdisciplinary network that extends beyond traditional PA research communities.

Two distinct representation effects may be revealed. First, the representation mechanism affects conceptual emphasis: author-supplied keywords foreground AI methods and technologies, whereas text-derived n-grams capture the broader governance, implementation, and application contexts in which AI is discussed. Second, the underlying corpus influences thematic orientation. Despite applying the same n-gram extraction method, the WoS and OpenAlex representations produce different conceptual structures. The WoS network is centered on governance and PA, whereas the OpenAlex network places greater emphasis on professional practice, education, ethics, and broader societal applications. These findings jointly demonstrate that alternative systems of scholarly representation do not merely identify different subsets of AI-in-PA literature; they construct different conceptual structures, disciplinary boundaries, and intellectual identities for the field.

**Discussion: Classification, Representation, and Boundaries**

The findings demonstrate that alternative systems of scholarly representation function more than identify different quantities of AI-in-PA scholarship. Across corpus size, publication types, publishing outlets, DOI overlap, temporal development, and conceptual structure and clustering, the representations frequently identify different publications, scholarly communities, and historical trajectories. AI-in-PA therefore does not emerge as a single, rigid, static body of literature to be discovered. Rather, its visibility and perceived boundaries depend, at least in part, on how scholarly knowledge is classified, retrieved, and represented.

The first implication concerns the scholarly communities through which AI-in-PA is communicated. The WoS Author representation is concentrated in established PA, public policy, and digital government venues, reflecting the traditional disciplinary core of the field. The OA Author representation, however, identifies a substantial body of AI-related scholarship published in social work journals, suggesting that important public-sector applications of AI are communicated to different professional audiences. The OpenAlex AI-assisted representations extend further into repositories, technical journals, and interdisciplinary publication platforms. These findings indicate that alternative representations not only retrieve different publications, but they are also grounded in distinct scholarships and professional communications.

Another implication centers on the relationship between disciplinary boundaries and systems of scholarly classification. OpenAlex is designed as a general-purpose knowledge organization system that classifies scholarship across the scientific literature using publication-level, AI-assisted methods. Such an approach is well suited for identifying broad conceptual relationships across disciplines. PA, however, has historically been defined by its object of study rather than by rigid disciplinary boundaries. Consequently, interdisciplinary AI-in-PA scholarship may be classified into neighboring fields such as computer science, information science, management, or social work rather than PA itself. This misalignment may help explain the paradox: OpenAlex provides broader coverage of PA scholarship overall, but it identifies substantially smaller AI-in-PA corpora than the WoS Author representation. A wider and more inclusive bibliographic coverage does not necessarily translate into greater discoverability or visibility for interdisciplinary domains or subfields.

The findings also underscore the continuing importance of human judgment in scholarly knowledge organization. Because systems of scholarly classification are not neutral but embody particular assumptions about disciplinary structure, conceptual similarity, and intellectual boundaries, AI-assisted representations should be interpreted as one perspective rather than an objective description of a field. Disciplinary boundaries and identities are not static but are continually negotiated and socially constructed through scholarly communities, publication practices, professional associations, citation networks, and evolving research agendas. Consequently, determining what belongs within a field is not solely a technical classification task but also an interpretive scholarly judgment. Especially for interdisciplinary domains such as AI-in-PA, expert disciplinary knowledge remains essential for evaluating whether algorithmic classifications adequately reflect the field's intellectual communities, publication practices, and evolving boundaries. Rather than replacing human expertise, AI-assisted knowledge organization is likely to be most valuable when combined with informed disciplinary judgment that can interpret, validate, and, when necessary, challenge algorithmically generated representations.

Several limitations warrant consideration. First, its broad scope captures AI-related PA scholarship across methodological, substantive, and applied contexts without systematically distinguishing among them. The findings therefore reflect the conceptual organization of the literature rather than the extent of AI adoption in research or practice. Second, differences in database coverage, disciplinary classification, textual representation, and DOI availability complicate direct comparisons between WoS and OpenAlex. Because the analysis uses both author-defined keywords and text-derived n-grams, the observed differences reflect alternative representations of the literature rather than the independent effects of either database. DOI-based overlap comparisons are also limited to records with identifiable DOIs and may not fully capture relationships among the complete corpora. The restriction to English-language publications, selected document types, and frequently occurring terms may also affect the coverage and thematic structure of the literature. Finally, average publication years indicate the relative timing of research themes but do not establish distinct developmental phases or turning points.

## Conclusions and Future Directions

This study examined whether alternative systems of scholarly classification identify and represent AI-in-PA scholarship similarly or differently. AI-in-PA provides a useful test case because it occupies an interdisciplinary domain with fluid and evolving boundaries. Drawing on five alternative representations derived from WoS and OpenAlex, including author-defined, citation-driven, and AI-assisted approaches, the study compared how different systems identify, organize, and characterize AI-in-PA scholarship.

The findings demonstrate that representation matters. The five representations frequently identified different bodies of knowledge rather than merely larger or smaller subsets of the same literature, positioned AI-in-PA within different scholarly and professional communities, and produced different historical and conceptual narratives of the field. These differences are reflected in the very limited overlap in publications and publishing outlets across the representations. Depending on the representation employed, AI-in-PA appears either as the gradual diffusion of externally developed, AI-related concepts into PA through interdisciplinary scholarship or as the recent emergence of a distinct area of PA research. These contrasting narratives reflect differences in scholarly representation rather than definitive evidence of the field's historical development. Likewise, the conceptual analyses show different emphases on AI methods, governance and administration, and professional and societal contexts. These findings suggest that alternative systems of scholarly representation influence not only what scholarship becomes visible but also how the field itself is understood. For bibliometric research in PA, these findings support greater transparency about disciplinary classification rules, the use of complementary retrieval strategies where appropriate, and explicit acknowledgment of the scholarship that particular approaches may exclude.

The broader contribution of this study extends beyond AI and PA. It demonstrates that systems of scholarly classification and algorithmic knowledge organization are not neutral infrastructures for organizing scientific information. Rather, they participate in representing disciplinary boundaries, scholarly communities, and intellectual identities. The findings also suggest that representation itself should become an explicit consideration in bibliometric research. As AI-assisted systems increasingly mediate scholarly discovery, evidence synthesis, research evaluation, and scientific mapping, questions of classification, representation, and knowledge organization become increasingly important. For PA scholars, this calls for greater attention to how database selection and classification approaches shape the literature used to examine emerging administrative problems and interdisciplinary research domains. Future research should examine whether similar representation effects occur in other interdisciplinary domains, including nonprofit studies, criminal justice, urban studies, public policy, and sustainability, where disciplinary boundaries are similarly fluid and contested. Further research could also assess the substantive relevance of publications identified by different classification approaches and examine how these representations change over time. Such work would help determine whether the patterns observed here are unique to AI-in-PA or reflect broader challenges in representing interdisciplinary scholarship.

## References

Bianchini, S., Di Girolamo, V., Ravet, J., & Arranz, D. (2026). AI in science: When and where it makes a difference. *Research Policy, 55*(6), 105478. https://doi.org/10.1016/j.respol.2026.105478

Chavarro, D., Ràfols, I., & Tang, P. (2018). To what extent is inclusion in the Web of Science an indicator of journal 'quality'? *Research Evaluation, 27*(2), 106–118. https://doi.org/10.1093/reseval/rvy001

Chawla, D. S. (2022, January 24). Massive open index of scholarly papers launches. *Nature*. https://doi.org/10.1038/d41586-022-00138-y

Chen, Z., & Schintler, L. A. (2023). Rediscovering regional science: Positioning the field's evolving location in science and society. *Journal of Regional Science, 63*(3), 617–642.

Frank, H., & Kunz, K. (2025). The need to reconnect public administration education, research, and practice. *Administration & Society, 57*(2), 310–335. https://doi.org/10.1177/00953997241283706

Frederickson, H. G. (1999). The repositioning of American public administration. *PS: Political Science & Politics, 32*(4), 701–712.

Gao, J., & Wang, D. (2024). Quantifying the use and potential benefits of artificial intelligence in scientific research. *Nature Human Behaviour, 8*, 2281–2292. https://doi.org/10.1038/s41562-024-02020-5

Goodnow, F. J. (1900). *Politics and administration: A study in government*. Macmillan.

Han, Y., Xiong, M., & Frank, H. A. (2020). Public administration and macroeconomic issues: Is this the time for a marriage proposal? *Administration & Society, 52*(9), 1439–1462. https://doi.org/10.1177/0095399720915292

Hao, Q., Xu, F., Li, Y., & Evans, J. (2026). Artificial intelligence tools expand scientists' impact but contract science's focus. *Nature, 649*, 1237–1243. https://doi.org/10.1038/s41586-025-09922-y

Kettl, D. F. (2000). Public administration at the millennium: The state of the field. *Journal of Public Administration Research and Theory, 10*(1), 7–34.

Li, B., Lyon, M. A., Ivonchyk, M., Appe, S., Dodge, J., Fox, A., Gil-Garcia, J. R., Rubin, E., & Luna-Reyes, L. F. (2025). Using generative AI in public affairs education and scholarly activity: A much-needed reflection. *Journal of Public Affairs Education, 31*(1), 9–17. https://doi.org/10.1080/15236803.2024.2429195

McDonald, B. D., III, Hall, J. L., O'Flynn, J., & van Thiel, S. (2022). The future of public administration research: An editor's perspective. *Public Administration, 100*(1), 59–71.

Ni, C., Sugimoto, C. R., & Robbin, A. (2017). Examining the evolution of the field of public administration through a bibliometric analysis of *Public Administration Review*. *Public Administration Review, 77*(4), 496–509. https://doi.org/10.1111/puar.12737

Olsen, A. L., Bendtsen, K.-E., & van Leeuwen, P. (2026). Public administration beyond public administration journals. *PS: Political Science & Politics, 59*(3), 501–507. https://doi.org/10.1017/S1049096526101942

Raadschelders, J. C. N. (2011). *Public administration: The interdisciplinary study of government*. Oxford University Press.

Riccucci, N. (2010). *Public administration: Traditions of inquiry and philosophies of knowledge*. Georgetown University Press.

Shark, A. R. (2024). *AI: A primer for state and local governments*. Public Technology Institute.

Vogel, R. (2014). What happened to the public organization? A bibliometric analysis of public administration and organization studies. *The American Review of Public Administration, 44*(4), 383–408. https://doi.org/10.1177/0275074012470867

Wagner, C. S., & Raadschelders, J. C. N. (2025). From disciplinary depth to interdisciplinary breadth: The case of public administration. *The American Review of Public Administration, 55*(4), 299–317. https://doi.org/10.1177/02750740241303106

Wang, C., Yin, Y., & Hu, H. (2026). The rise of algorithmic governance and the dual revolution: Applications, challenges, and governance of artificial intelligence in public administration. *Technology in Society, 86*, 103264. https://doi.org/10.1016/j.techsoc.2026.103264

Wilson, W. (1887). The study of administration. *Political Science Quarterly, 2*(2), 197–222.

Zhu, X. (2026). Public administration with, of, and through AI: Toward a new paradigm in the era of intelligence. *Journal of Chinese Governance, 11*(1), 30–57. https://doi.org/10.1080/23812346.2025.2578589

**Table 1: Layered Framework of AI-Related Search Keywords**

| Layer | Search Keywords | Explanation |
|---|---|---|
| Foundation | artificial intelligence (AI); machine learning (ML); representation learning | This layer captures the methodological foundations of AI, focusing on data-driven learning approaches such as machine learning and representation learning. It reflects how AI systems learn patterns, relationships, and abstract features directly from data through optimization and statistical inference |
| Architecture | deep learning (DL); natural language processing (NLP; nlp); multimodal AI | This layer focuses on how AI systems are built and what kinds of information they can understand. Deep learning provides neural network-based architectures, and NLP and multimodal AI extend these capabilities to language and multiple data types, e.g., text, images, audio |
| Model | foundation model (FM); large language model (LLM); generative pre-trained transformer (GPT) | This layer refers to large-scale, pre-trained models that serve as general-purpose computational engines. All can be adapted to many different tasks without being redesigned from scratch |
| Action | generative AI (GenAI); agentic AI; AI agent | This layer captures the operational capabilities of AI systems, focusing on what they do in practice. This includes generating content, e.g., text, images, or code, and increasingly acting autonomously through agents that can make decisions, use tools, and execute multi-step tasks |

**Table 2: AI-in-PA Identification Approaches**

| Approach | Database | Search Fields | Classification Logic | Description |
|---|---|---|---|---|
| **WoS Author** | Web of Science | Titles, Abstracts, Author-Supplied Keywords | Author-defined | Identifies AI-in-PA scholarship using authors' own descriptions of their research |
| **WoS Keywords Plus** | Web of Science | Keywords Plus | Citation-driven | Identifies AI-in-PA scholarship using algorithmically generated terms derived from cited references |
| **OA Author** | OpenAlex | Titles, Abstracts | Author-defined | Identifies AI-in-PA scholarship using authors' titles and abstracts. Author-supplied keywords are not available in OpenAlex; OpenAlex Keywords are algorithmically generated and therefore analyzed separately |
| **OA Concepts** | OpenAlex | Concepts | AI-assisted | Identifies AI-in-PA scholarship using algorithmically assigned OpenAlex concepts |
| **OA Keywords** | OpenAlex | Keywords | AI-assisted | Identifies AI-in-PA scholarship using algorithmically generated OpenAlex keywords |

**Table 3: DOI Availability and Overlap Across AI-in-PA Representations**

| Approach | Total Records | Unique DOIs | DOI Share (%) | WoS Author (DOIs) | WoS Keywords Plus (DOIs) | OA Author (DOIs) | OA Concepts (DOIs) | OA Keywords (DOIs) |
|---|---|---|---|---|---|---|---|---|
| **WoS Author** | 959 | **923** | 96.3 | **923** | 79 | 8 | 2 | 2 |
| **WoS Keywords Plus** | 133 | **125** | 94.0 | 79 | **125** | 1 | 0 | 0 |
| **OA Author** | 204 | **189** | 92.7 | 8 | 1 | **189** | 34 | 19 |
| **OA Concepts** | 494 | **282** | 57.1 | 2 | 0 | 34 | **282** | 235 |
| **OA Keywords** | 387 | **235** | 60.7 | 2 | 0 | 19 | 235 | **235** |

Note: DOI share is calculated as the number of unique DOI-linked publications divided by the total records within each approach. Off-diagonal values indicate the number of publications sharing the same DOI across two approaches, while diagonal values indicate the number of unique DOI-linked publications within each approach. Overlap comparisons exclude records without identifiable DOIs and therefore may not fully reflect overlap among the complete corpora, particularly for OA Concepts and OA Keywords.

**Table 4: Temporal Profiles of Alternative AI-in-PA Representations**

| Approach | Min Year | Max Year | Mean Year/Month | Median Year | Total Records |
|---|---|---|---|---|---|
| **WoS Author** | 1989 | 2026 | 2022.8 | 2024 | 959 |
| **WoS Keywords Plus** | 2019 | 2026 | 2024.1 | 2024 | 133 |
| **OA Author** | 1980 | 2026 | 2020.3 | 2023 | 204 |
| **OA Concepts** | 1980 | 2026 | 2010.5 | 2011 | 494 |
| **OA Keywords** | 1980 | 2026 | 2009.3 | 2012 | 387 |

Note: Minimum and maximum years indicate the earliest and latest publication years identified within each representation. Mean year is expressed as a decimal year, while median year represents the midpoint of the publication-year distribution. Total records indicate the number of publications identified by each approach. These statistics describe the temporal profiles of the retrieved corpora and should not be interpreted as definitive evidence of the historical emergence or development of AI-in-PA scholarship.

**Table 5: Top Ranked Terms Across Alternative Representations**

| Rank | WoS Author-Supplied Keywords | WoS Title/Abstract N-grams | OpenAlex Title/Abstract N-grams |
|---|---|---|---|
| 1 | Artificial Intelligence | Artificial Intelligence | Artificial Intelligence |
| 2 | Machine Learning | Public Sector | Policy |
| 3 | Natural Language Processing | Political | Practice |
| 4 | Public Sector | Machine Learning | Students |
| 5 | Governance | System | Education |
| 6 | Public Administration | Role | Digital |
| 7 | Generative AI | Regulatory | Professional |
| 8 | AI Governance | Impact | Social Workers |
| 9 | Regulation | Including | Ethical |
| 10 | Big Data | Issues | Technology |
| 11 | Digital Transformation | Legal | Future |
| 12 | E-Government | Accountability | Article |
| 13 | Social Media | Performance | Challenges |
| 14 | Deep Learning | Key | Role |
| 15 | Digital Government | Context | Political |
| 16 | Large Language Models | Understanding | Field |
| 17 | Public Policy | Public Administration | Critical |
| 18 | Accountability | Processes | Information |
| 19 | Sentiment Analysis | Related | Performance |
| 20 | Transparency | Application | Time |
| 21 | Digital Governance | Administrative | Impact |
| 22 | Open Data | Implications | Tools |
| 23 | COVID-19 | Identify | Knowledge |
| 24 | Ethics | Studies | Governance |
| 25 | Government | Effective | Government |
| 26 | AI Ethics | Address | Development |
| 27 | Twitter | Knowledge | Findings |
| 28 | China | Critical | Potential |
| 29 | European Union | Recent | Systems |
| 30 | Topic Modeling | Institutional | Implications |

Note: Colors indicate broad thematic categories.

AI Methods & Technologies: AI concepts, models, computational techniques, and analytical methods.

Governance, Policy & Administration: public administration, governance, regulation, government institutions, and public-sector processes.

Societal, Organizational & Professional Contexts: implementation, professional practice, education, organizations, users, impacts, and broader societal applications of AI.

**Figure 1: Publication Types of PA Records**

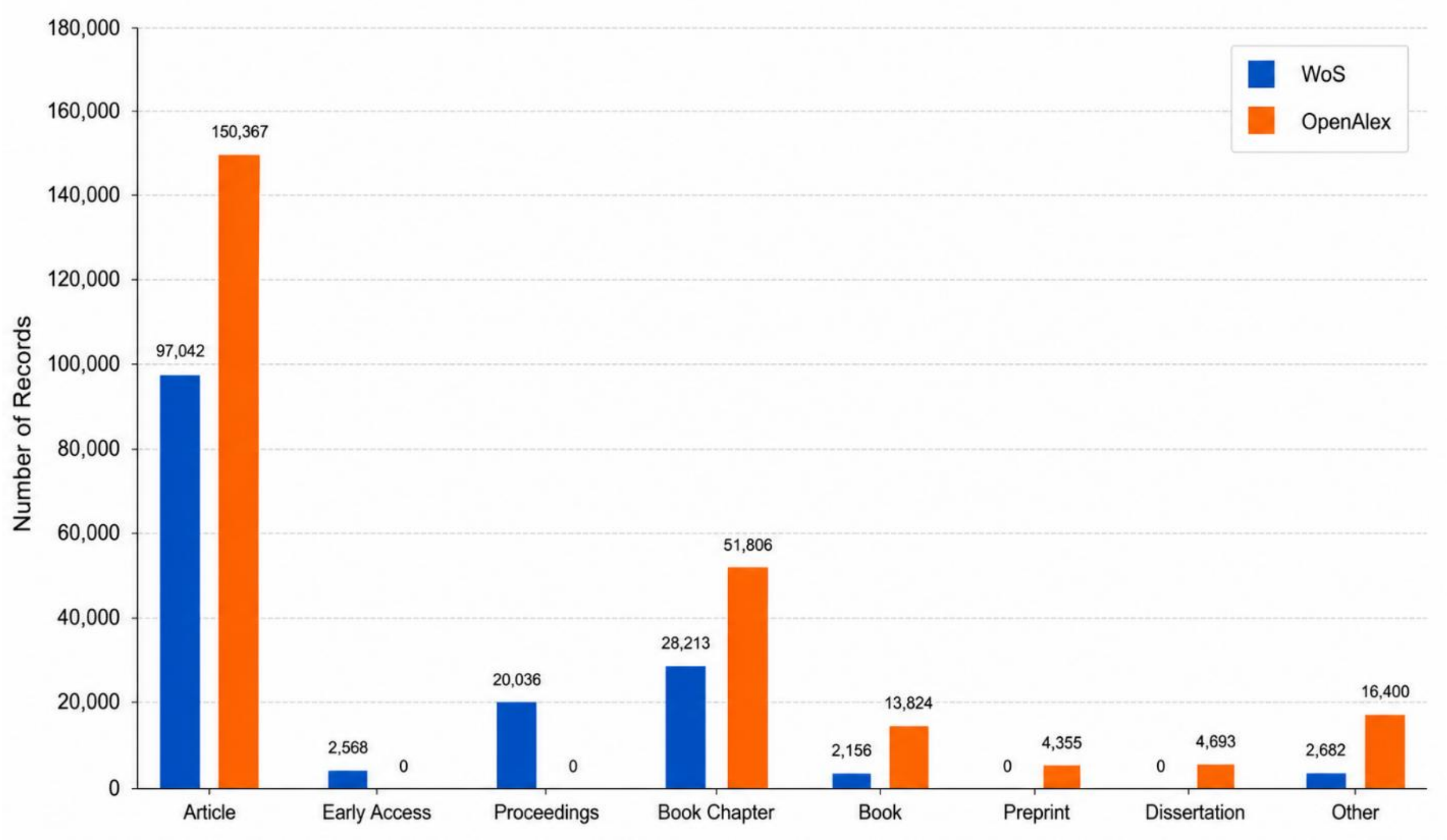

**Figure 2: AI-in-PA Publication Counts Across Five Representations**

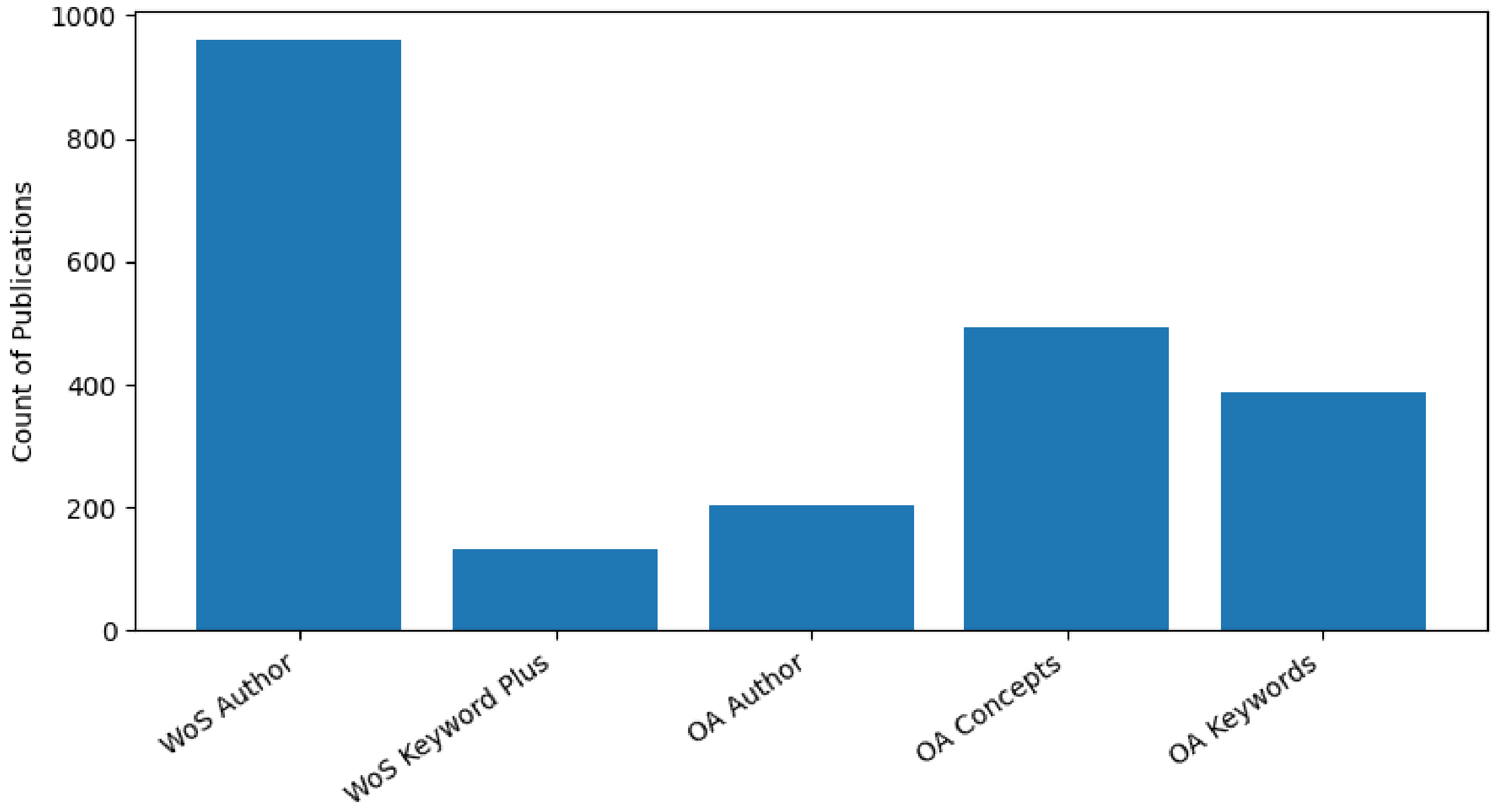

**Figure 3: Type of Publications in AI-in-PA Scholarship**

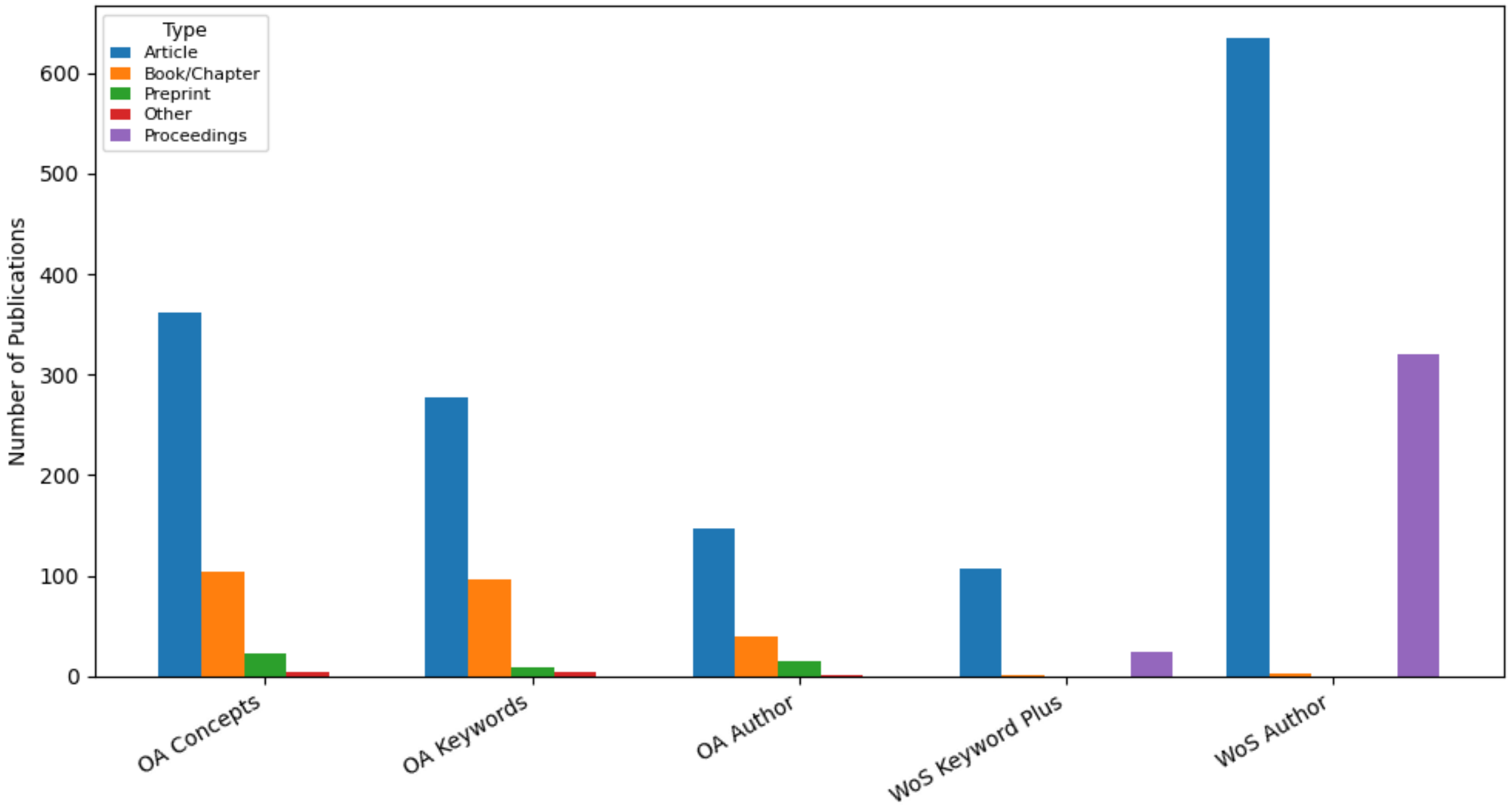

Note: Bars indicate the number of publications by document type within each AI-in-PA representation. Document types follow the respective classifications of WoS and OpenAlex and may not be directly comparable across databases.

**Figure 4: Overlapping of Top Publication Outlets**

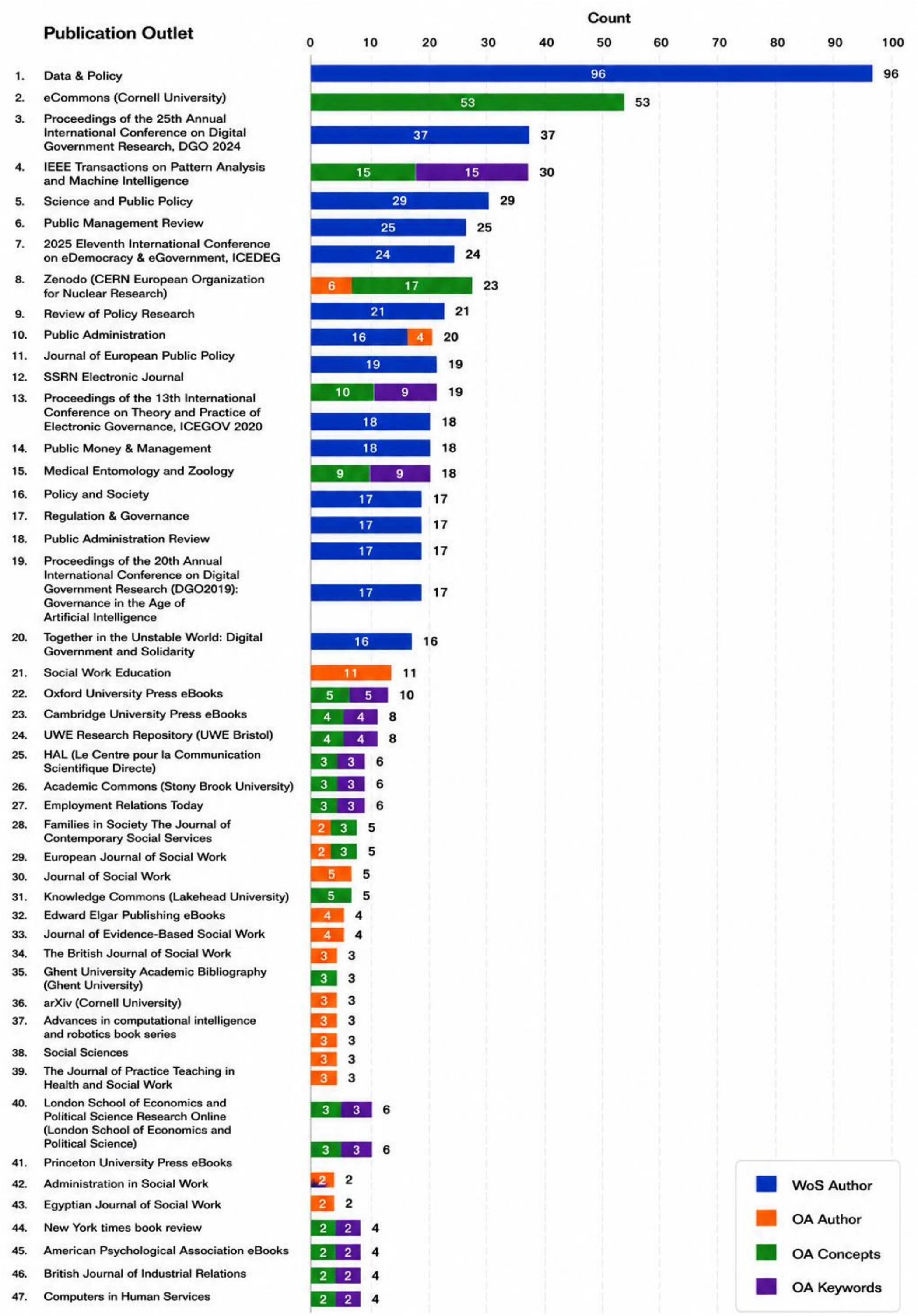

Note: Bars indicate the number of publications associated with each outlet across the four representations shown. Colors distinguish the representations, and stacked segments indicate their respective contributions to each outlet's total. Publication outlets include journals, conference proceedings, institutional repositories, and other scholarly sources, as recorded in WoS and OpenAlex.

**Figure 5: Overlap in Outlets and Publications Across Representations**

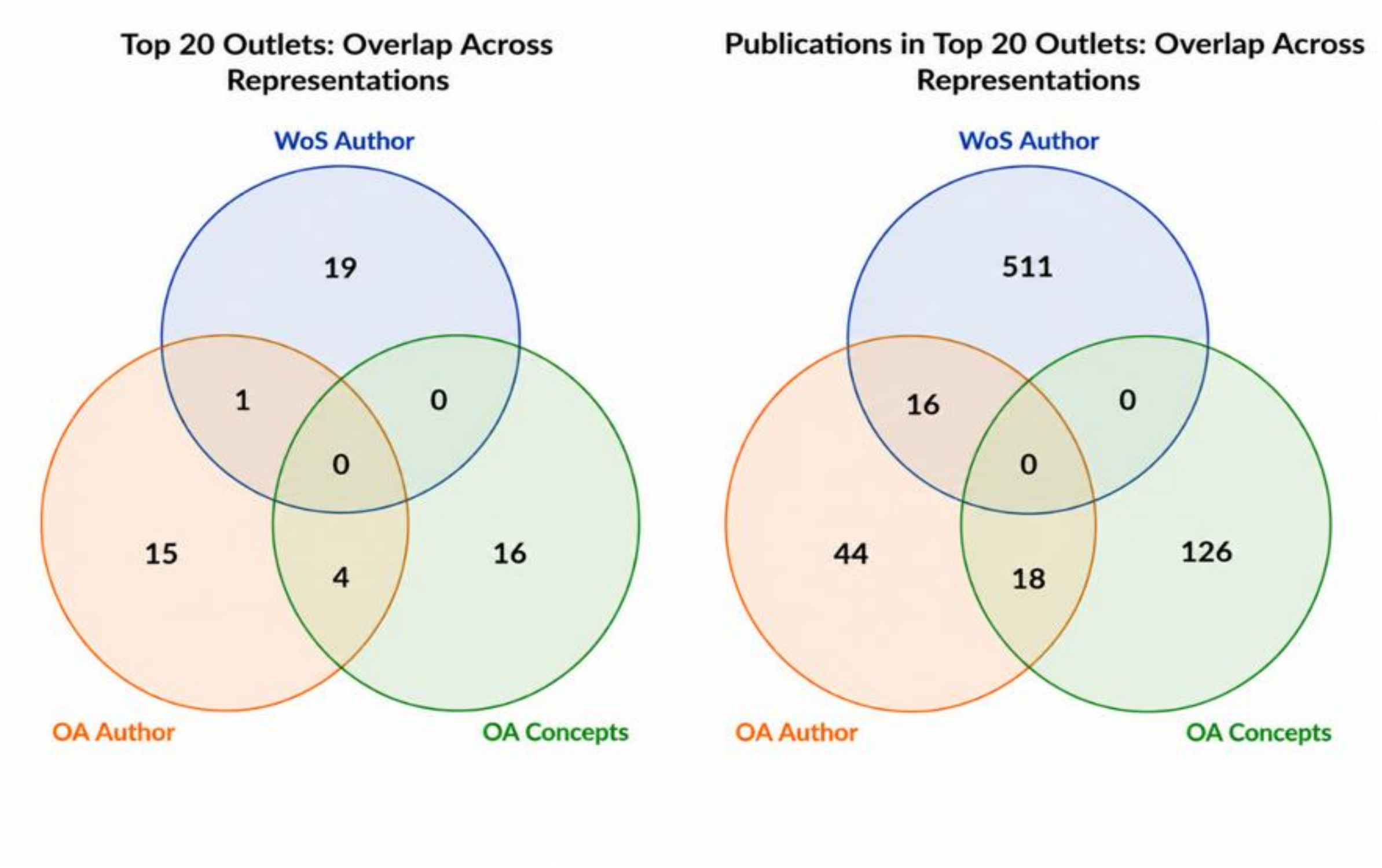

Note: The left panel shows overlap among the top 20 publication outlets identified separately by WoS Author, OA Author, and OA Concepts. The right panel shows the corresponding publication counts associated with these outlets. Values in overlapping regions indicate outlets or publications shared across the respective representations.

**Figure 6: Cumulative AI-in-PA Scholarship, 1980-2026**

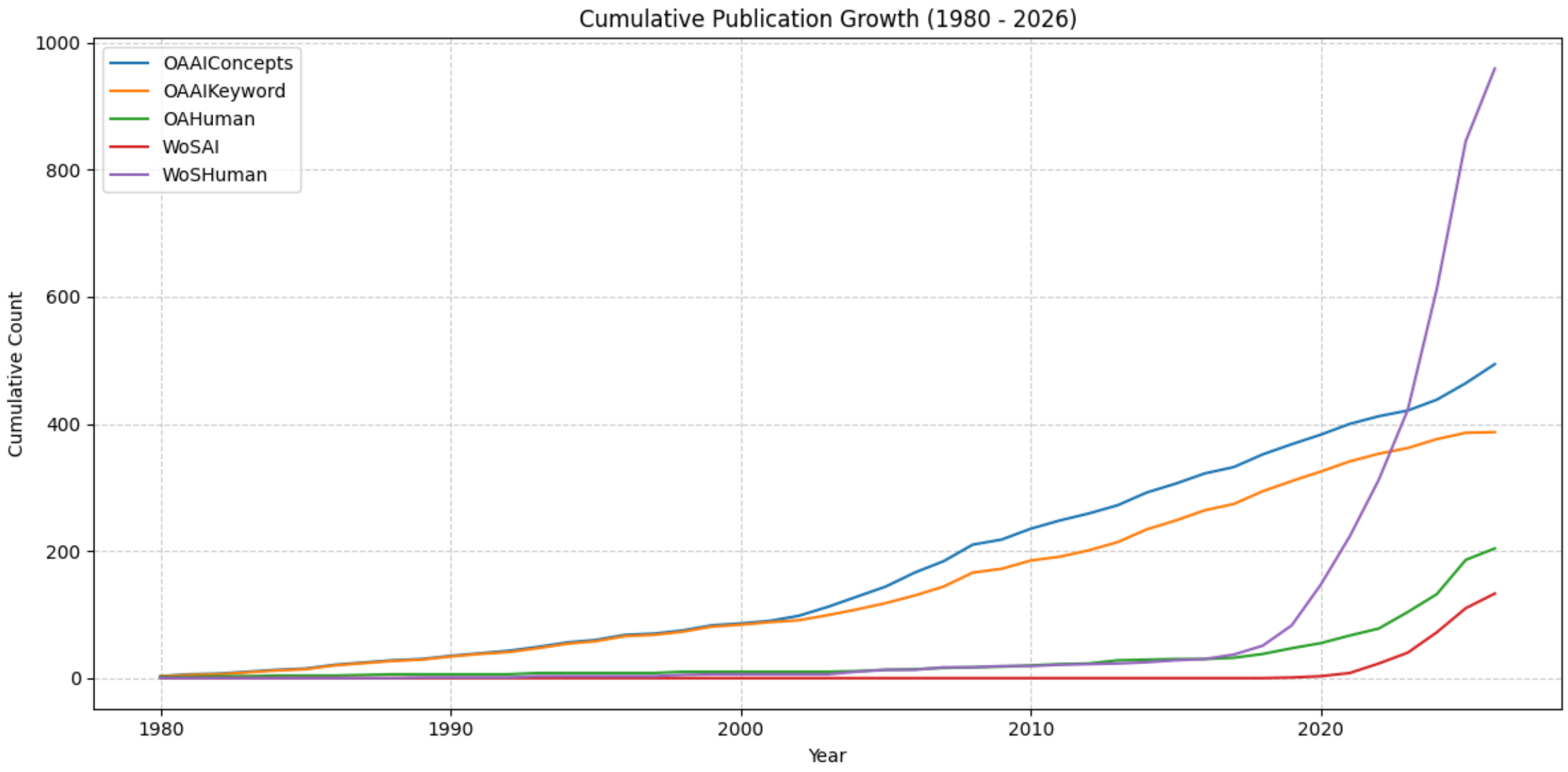

Note: Lines represent cumulative annual publication counts identified by each of the five AI-in-PA representations from 1980 to 2026. Counts are accumulated separately within each representation, and publications may appear in more than one corpus. The trajectories reflect differences in retrieval and classification rather than definitive patterns in the historical development of AI-in-PA scholarship.

**Figure 7: Thematic Clusters Across Alternative AI-in-PA Representations**

| WoS Author-Supplied Keywords | WoS Title and Abstract N-gram | OpenAlex Title and Abstract N-gram |
|---|---|---|

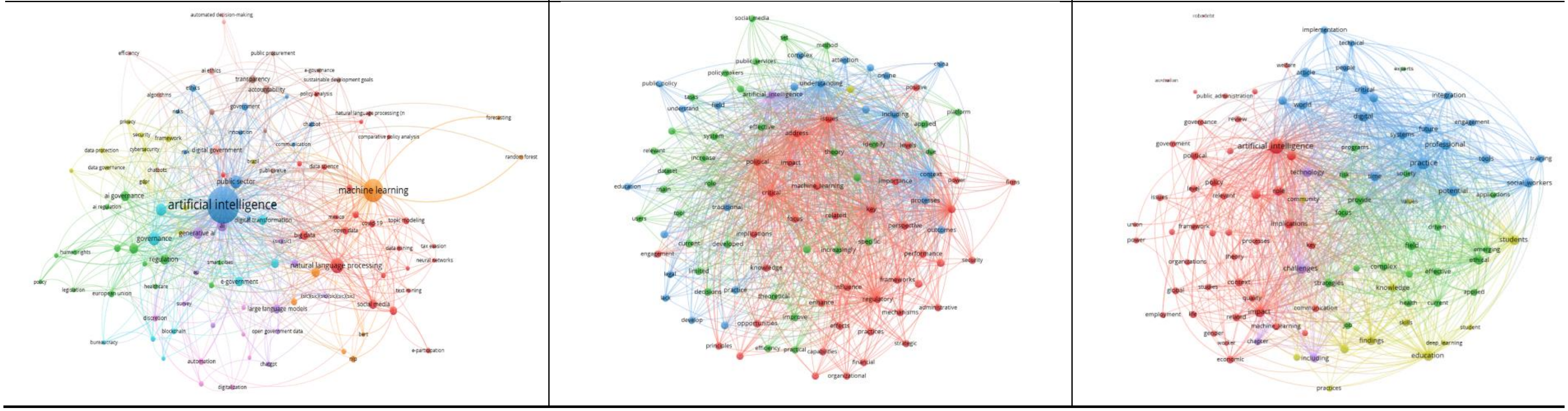

Note: Panels (a), (b), and (c) present term co-occurrence networks constructed from WoS author-supplied keywords, WoS title/abstract n-grams, and OpenAlex title/abstract n-grams, respectively. Nodes represent terms, with node size indicating term prominence and links representing document-level co-occurrence. Colors distinguish thematic communities identified through the Louvain algorithm. Because the networks are constructed separately, node sizes, spatial positions, and cluster colors are not directly comparable across panels.

**Appendix 1: Full Search String of Web of Science**

```
(
 (
   (TI=( "artificial intelligence" OR "machine learning" OR "representation learning" OR "deep learning" OR "natural language processing" OR "NLP" OR "multimodal AI" OR "foundation model" OR "generative AI" OR "GenAI" OR "agentic AI" OR "AI agent" ) OR
    AB=( "artificial intelligence" OR "machine learning" OR "representation learning" OR "deep learning" OR "natural language processing" OR "NLP" OR "multimodal AI" OR "foundation model" OR "generative AI" OR "GenAI" OR "agentic AI" OR "AI agent" ) OR
    AK=( "artificial intelligence" OR "machine learning" OR "representation learning" OR "deep learning" OR "natural language processing" OR "NLP" OR "multimodal AI" OR "foundation model" OR "generative AI" OR "GenAI" OR "agentic AI" OR "AI agent" ))
   AND PY=(1980-2026)
 )
 OR
 (
   (TI=("large language model" OR "LLM" OR "generative pre-trained transformer" OR "GPT") OR
    AB=("large language model" OR "LLM" OR "generative pre-trained transformer" OR "GPT") OR
    AK=("large language model" OR "LLM" OR "generative pre-trained transformer" OR "GPT"))
   AND PY=(2023-2026)
 )
)
AND WC=("Public Administration")
```

## Appendix 2: OpenAlex Retrieval and AI-Related Publication Identification

This appendix provides a procedural description of the OpenAlex retrieval and screening workflow to facilitate its replication. The procedure differs from the Web of Science (WoS) search in two important respects. First, PA scholarship is identified through OpenAlex's automated publication-level topic classification rather than the source-level classifications employed by WoS. Second, AI-related publications are identified through subsequent keyword-based lexical filtering in R rather than directly through the database search interface.

Bibliographic records were retrieved in April 2026 from the OpenAlex Works API (https://api.openalex.org/works) using custom R scripts. The retrieval covered publications from 1980 to 2026 whose primary topics were assigned to the Public Administration subfield (3321), using the classification field primary_topic.subfield.id. Retrieved metadata included publication identifiers, titles, publication years, document types, languages, author affiliations, classification information, and abstract inverted indices.

The retrieved records were subsequently processed in R. Abstract inverted indices were reconstructed into plain text by arranging words according to their recorded positions, and duplicate records were removed using unique OpenAlex identifiers (id). English-language publications were retained for subsequent screening. AI-related publications were then identified by searching publication titles and reconstructed abstracts for terminology corresponding to the four-layer conceptual framework presented in Table 1, encompassing foundation, architecture, model, and action. The lexical screening followed the AI-related search vocabulary employed in the WoS analysis (Appendix 1).

This two-stage approach combined OpenAlex's disciplinary classification with subsequent keyword-based screening to identify AI-related PA scholarship. The workflow provides a consistent conceptual basis for identifying relevant publications across the two databases while accommodating differences in their classification systems and retrieval procedures. Because OpenAlex is continuously updated, including changes to its bibliographic records, metadata, and topic classifications, repeating the same workflow at a later date may not reproduce the exact dataset retrieved in April 2026. This distinction between replicability of the retrieval procedure and reproducibility of its historical results reflects the evolving nature of OpenAlex rather than a limitation of the procedure itself.